\documentclass{aa}
\usepackage{graphicx}
\usepackage{natbib}
\begin{document}
\newcommand{\Msol}{M$_{\odot}$}
   \title{COMPTEL upper limits for the $^{56}$Co  $\gamma$-ray emission from SN1998bu}

   \author{R. Georgii\inst{1,2}
          \and
          S. Pl\"{u}schke\inst{1}
          \and
          R. Diehl\inst{1}
          \and
          G.G. Lichti\inst{1}
          \and
          V. Sch\"{o}nfelder \inst{1}
          \and
          H. Bloemen\inst{3}
          \and
          W.Hermsen\inst{3}
          \and
          J. Ryan\inst{4}
          \and
          K. Bennett\inst{5}
   }

   \offprints{R. Georgii, \email{rgeorgii@frm2.tum.de}}

   \institute{Max-Planck-Institut f\"{u}r extraterrestrische Physik, Postfach 1312,
              D-85741 Garching, Germany
         \and
             FRM-II, Technische Universit\"{a}t M\"{u}nchen,
              Lichtenbergstr. 1, D-85747 Garching, Germany
         \and
             SRON Institute for Space Research, Sorbonnelaan 2, 3584 CA Utrecht, The Netherlands
         \and
             Space Science Center, University of New Hampshire, Durham, NH 03824, USA
         \and
             Science Division, ESTEC, ESA, 2200 AG Noordwijk, The Netherlands
   }

   \date{Received 19 Feb 2002; accepted 1 Aug 2002}

   \abstract{
    Supernova 1998bu in the galaxy M96 was observed by COMPTEL for a total of 88 days starting
    17 days after the explosion. We searched for a signal in the 847 keV and 1238 keV lines of
    radioactive $^{56}$Co from this type Ia supernova. Using several different analysis methods, 
    we did not detect SN1998bu. Our measurements should have been sensitive enough to detect
	$^{60}$Co gamma-rays as predicted from supernova models. 
	Our $2\sigma$ flux limit is 
        $2.3 \cdot 10^{-5}$ photons~cm$^{-2}$~s$^{-1}$;
    this would correspond to 0.35 \Msol~of ejected $^{56}$Ni, if SN1998bu 
	were at a distance of 11.3 Mpc
	and transparent to MeV gamma rays for the period of our measurements.
	We discuss our measurements in the context of common supernova models, 
	and conclude disfavoring a supernova event with large mixing and major parts of the
    freshly-generated radioactivity in outer layers. 
   }
   \maketitle
%

\section{Introduction}
Despite their widespread use as 'standard candles', the physical
nature of supernovae of type Ia is still not understood in terms 
of physical processes; therefore
corrections of evolutionary aspects remain empirical 
\citep{Georgii:Branch1998,Georgii:Niemeyer2000}.
Type Ia supernovae are believed to be caused by thermonuclear explosions of 
CO white dwarfs \citep{Georgii:Livio2000,Georgii:Hoeflich1996,Georgii:Nomoto97}. 
Radioactive energy of $\simeq$0.5 \Msol\ of $^{56}$Ni 
synthesized in such explosions is considered the driver of all types of observed light
from these objects.
The nearby SN1998bu supernova was a unique opportunity to directly measure
gamma-rays from the $^{56}$Ni decay chain with the Compton Observatory.

The dynamics of a white dwarf explosion are
difficult to model due to the range of scales
involved in flame ignition and propagation \citep{Georgii:Iwamoto1999}. 
Theories which are most successful in describing observations of type Ia supernovae include 
empirical components for key aspects. 
Observationally type Ia supernovae are a fairly homogeneous phenomenon \citep{Georgii:Branch1998},
suggesting a narror range of synthesized $^{56}$Ni masses. 
Constraints from  models of the bolometric light curve of type Ia supernovae
imply that typically 0.3--0.5 \Msol\ of radioactive $^{56}$Ni energy are needed\citep{Georgii:Hoeflich1996}.
Observations of the supernova light curve's 
peak magnitude or of NIR lines of Fe[II] and Co[II] have mostly been used to derive $^{56}$Ni masses; 
the rather wide range of  inferred $^{56}$Ni (0.1 -- 1.14 \Msol)\citep{Georgii:Contardo2000} is difficult
to understand if a single well-tuned process is held responsible for the supernovae of type Ia and
in particular their "standard candle" characteristics. 
Among different types of models, a rather wide range of $^{56}$Ni masses of 
0.1 up to 0.8 \Msol\ is discussed (see `Discussion' section below). 

Critical parameters of the explosion models are the ignition density of the white dwarf at its core, and the
transition from the early deflagration stage (sub-sonic flame propagation) into a detonation
(super-sonic flame propagation)\citep[e.g.][]{Georgii:Leibundgut2000,Georgii:Livio2000}. 
The former is estimated from evolutionary models of
the (uncertain) progenitor and the effective binary accretion rate, involving uncertain issues
such as steady or flash-like nuclear burning of the accreted H and He material, or modulation of
the accretion flow from the companion through the wind of the white dwarf (at the higher 
accretion rates required by the lower ignition densities preferred from otherwise excessive
production of neutron-rich Fe group isotopes\citep{Georgii:Nomoto97}). 
The acceleration of the flame speed can only
be treated as a purely empirical parameter of models at present, but critically
determines the final $^{56}$Ni mass and the Fe group to lighter element ratio
\citep{Georgii:Iwamoto1999}.
Three-dimensional model treatments of this flame
``micro-physics'' is promising, but still a challenging problem \citep{Georgii:Reinecke1999}.

The fact that
type Ia supernovae are rather homogeneous \citep{Georgii:Branch1998}
suggests a clear evolutionary path towards a well-definied presupernova star and a robust ignition condition. 
This led to the 
model of binary accretion of H and He rich matter onto a CO white dwarf at a well-tuned
accretion rate such that H and especially He nuclear burning proceeds 
non-catastrophically and the white dwarf C mantle grows in mass
until the Chandrasekhar mass limit is reached; the thermonuclear runaway explosion ensues
from fast nuclear burning of Carbon ignited at the core due to heating from gravitational pressure 
and H and He shell-burning heat conduction \citep{Georgii:Nomoto1982}.
Sub-Chandrasekhar mass white dwarfs could also explode as type Ia supernovae \citep{Georgii:Livne1990}:
The merging of two white dwarfs would disrupt the lighter of the two into a C-O envelope accreting
onto the more massive white dwarf ("double-degenerate" model). As the accretion proceeds to exceed
the Chandrasekhar limit, the white dwarf ignites C centrally as above \citep{Georgii:Iben1984}.
There are some doubts if the merging process will avoid core collapse of the merged object and produce
a thermonuclear supernova; e.g. transport of rotational energy is critical \citep{Georgii:Livio2000}.
Alternatively, a single-degenerate sub-Chandrasekhar model has been proposed:
A He layer built up from accretion and steady hydrogen burning as above may ignite in a flash
and thus send a shock wave into the white dwarf core, adding to the gravitational heat and thus
also igniting carbon in the center for a lower-mass white dwarf \citep{Georgii:Nomoto1982,Georgii:Livne1995}.
In this scenario much of the radioactive material would be produced 
towards the outside, resulting in different evolution of radioactivity-derived
supernova light. Recent constraints on early spectra and the absence of intermediate-mass elements
in the outer fast ejecta disfavor this scenario somewhat \citep{Georgii:Livio2000}.

A deflagration model was 
recently favored for SN1998bu on purely spectroscopic 
arguments \citep{Georgii:Vinko2001}. And 
the occurrence near or in a spiral arm and the observation of a light-echo \citep{Georgii:Cappellaro2001} 
may suggest that the progenitor system could be younger than those of the average type Ia supernovae, 
therefore anomalous and igniting at a particularly low density.

With such diversity of models and the difficulties
of detailed physics modeling, observations of a variety of aspects of type Ia supernovae are 
a key to clarify the true nature of these events.
   
Observations of a large sample of supernovae in UV, optical and infrared bands
have been made and discussed widely. But this radiation
originates from driving processes deep inside the object, the bolometric
light curve and its evolution reflect the supernova envelope structure,
with much less information on the core. Spectral information tells us about 
material mixing and the total kinetic energy. However optical photons are created
long after the initial explosion; most information
from the early stage of the supernova event is lost.
This makes it difficult to discriminate between different explosion models or model parameters.

\begin{figure*}[ht]
\centering
\begin{minipage}[t]{8.5cm}
\includegraphics[width=\textwidth,bb= 5 8 185 190]{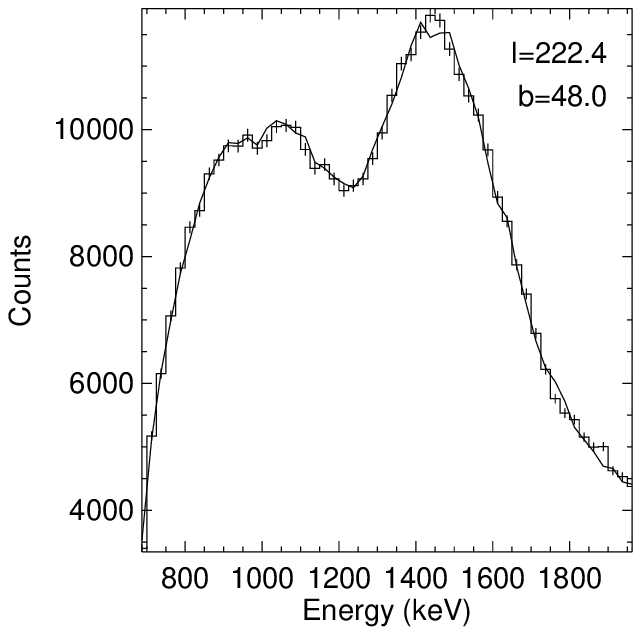}
\end{minipage}
\hfill
\begin{minipage}[t]{8.5cm}
\includegraphics[width=\textwidth, bb= 5 8 185 190]{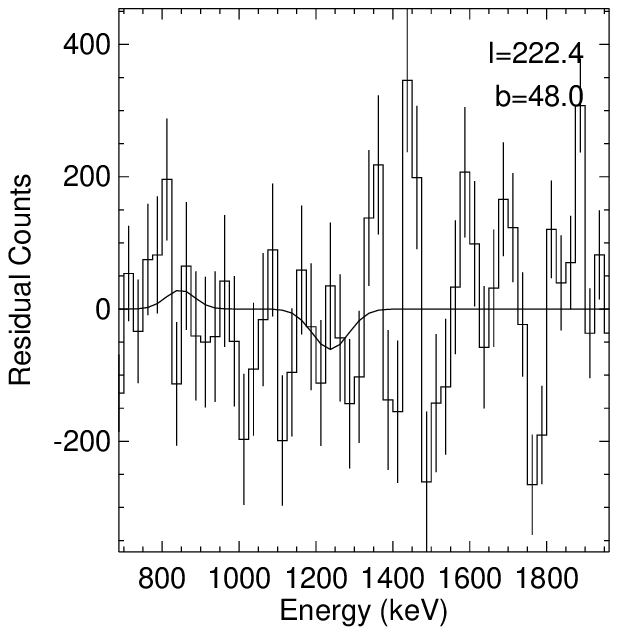}
\end{minipage}
\caption{The total energy spectrum (left) and the background subtracted spectrum (right) for a sample position on the grid around SN1998bu.}
\label{Georgii:F1}
\end{figure*} 

Observations of  $\gamma$-rays promise more direct information from 
the core of the supernova and the explosion mechanism.
The radioactivity produced in the initial event decays and produces
gamma-ray lines, which can be observed directly once the supernova is
transparent to gamma-rays (after about 30-100 days). Even in the early stages, $\gamma$-rays
from radioactivity will escape from outer layers, their intensity depending on ejecta mixing.
Differences in the predicted $\gamma$-ray spectra have been suggested as the key observation to
discriminate between models and the extent of mixing \citep{Georgii:Hoeflich1998,Georgii:Pinto2001}. 
This is best observed at early times; at times later
than ~30 days, differences in total $^{56}$Ni masses can mimic differences
between sub- and Chandrasekhar models \citep{Georgii:Pinto2001}.

The sensitivity of the  $\gamma$-ray instruments on-board CGRO
(OSSE and COMPTEL) limits the observations of type Ia
supernovae to events  within about 15 Mpc. 
To date, only one event, SN1991T, was marginally detected
with COMPTEL \citep{Georgii:Morris1997}. SN1998bu provides a second
opportunity for line searches in { type Ia supernovae}. 
Some theoretical models predict  $\gamma$-ray line
fluxes well above the sensitivity limits of COMPTEL and OSSE for
an assumed distance of 11~Mpc. Although COMPTEL lacks the spectral resolution
to provide unique and decisive $\gamma$-ray line shape diagnostics,
an independent proof of the radioactive $^{56}$Ni mass origin through detection
of the corresponding $\gamma$-ray line fluxes was attempted, and is important given
the complexity and unknowns of conversion of radioactive energy in a supernova envelope.

\begin{figure*}[tb]
\sidecaption
\includegraphics[width=12cm,,clip]{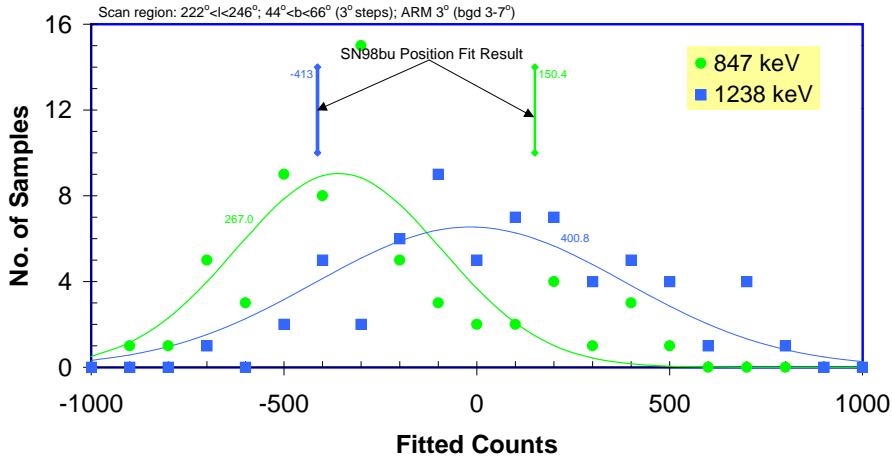}
\caption{The result of the spectral analysis. The derived $2\sigma$ upper limits are  
$4.1 \cdot 10^{-5}$ photons~cm$^{-2}$~s$^{-1}$ for the 847 keV line and 
$2.3 \cdot 10^{-5}$ photons~cm$^{-2}$~s$^{-1}$ 
for the 1238 keV line.}
\label{Georgii:Line}
\end{figure*}

On May 9.9 UT in 1998 supernova SN1998bu was discovered in the galaxy M96 (NGC 3368)
\citep{Georgii:Villi98}. From wide-band spectrograms it was classified as type Ia
\citep{Georgii:Ayani98,Georgii:Meikle1998}. From an earlier
observation \citep{Georgii:Faranda1998} and an { estimate of
the} maximum blue light { at} $t_{\rm Bmax} =$ 10952.7 $\pm$ 0.5
TJD  (i.e.~May 19), \cite{Georgii:Meikle1998} inferred the
date of the explosion to be May 2.0 $\pm$ 1.0 UT (i.e. TJD 10935
$\pm$ 1). The distance to M96 had been known from HST Cepheid measurements as 
11.3 $\pm$ 0.9~Mpc \citep{Georgii:Hjorth1997}, though another value of 9.6 $\pm$ 0.6~Mpc had been 
derived from measurements of planetary nebulae \citep{Georgii:Feldmeier1997}.
SN1998bu appears to be a typical type Ia event \citep{Georgii:Jah1999}, its reddening can be
attributed to dust in the host galaxy. From its dereddened brightness it was concluded (using the methods described
in \citet{Georgii:Nomoto97} and \citet{Georgii:Iwamoto1999}) that the 
total ejected Ni mass was rather typical. A value of 0.77 \Msol\ has been derived 
from analysis of the bolometric light curve \citep{Georgii:Leibundgut2000,Georgii:Contardo2000}.
 
\section{Observations}
The COMPTEL observations of SN1998bu began on TJD 10952, 17 days
after the explosion. Due to this late start and
to the low sensitivity of COMPTEL for low energies, we missed the opportunity to measure
the decay { lines of $^{56}$Ni ($\tau = 8.8\,$d) at  750~keV
and 812~keV.} { $^{56}$Co, the daughter nucleus of $^{56}$Ni,
has a mean life of $\tau = 112\,$d and decays to stable 
$^{56}$Fe. In this decay two  $\gamma$-ray lines are emitted at 847~keV
and 1238~keV. The COMPTEL observations spanned a total duration of 88 days, 
ending at TJD 11071, 136 days after the explosion.}

The COMPTEL telescope detects $\gamma$-rays through a Compton scatter interaction
in its upper plane of liquid-scintillation detectors (``D1''), followed by
detection of another interaction in the lower plane of NaI(Tl) scintillation
detectors (``D2''), ideally totally absorbing the scattered $\gamma$-ray.
Imaging information is provided by the Compton scatter kinematics through
energy deposit and interaction location measurements
\citep[For a detailed description of COMPTEL
see][]{Georgii:Schoenfelder1993}.

\section{Analysis and Results}
 To obtain a higher sensitivity at lower energies the
observations were { carried out} in  a ``low-threshold'' mode,
where the { energy thresholds} of the D2 modules were
reduced below the normal 600 -- 700 keV range; the normal mode would have
degraded the instrument response for the 847 keV photons.
Two different analysis approaches were used: "spectral" and "imaging" analysis.

\subsection{Spectral analysis}
In this approach the full spectral response, i.e.~the photo peak and the Compton tail, 
is exploited. Imaging parameters of the measurement are used for coarse field-of-view 
selection of the events. Only events with D2 energy deposits above 600 keV were used to avoid the 
511 keV line background which originates from positrons. These result mainly from $^{22}$Na within 
the instrument,
one of the major sources of instrumental background, which  has its origin in activation of COMPTEL 
passing through the radiation belts
\citep{Georgii:Weidenspointner2001}.

Spectra were derived for a $\sim 12^{\rm o}$ FWHM beam on a grid around the SN position
with a grid spacing of $3^{\rm o}$. Fig.~\ref{Georgii:F1} shows an example for the 
direction covering the SN position. 
The beam size was chosen through the selections of events from within a $2^{\rm o}$ wide range around the 
pivot direction, performed near the cone-shaped response in the COMPTEL data space.
Background spectra were constructed from the 
same direction, selecting a ring-like $3^{\rm o}-7^{\rm o}$ wide region around the
same pivot direction. 
Background-subtracted spectra were then analyzed for the relevant Co decay lines 
(see  Fig.~\ref{Georgii:F1}). Gaussian-shaped lines with instrumental resolution
are fitted to the data to determine line intensities. 
This analysis and in particular the background determination may include unknown 
systematic effects that  could mimic lines. To account for these uncertainties
(in addition to the statistical uncertainties), we determined the variance of the
fitted line intensities across our measured sample empirically:
We histogrammed the fitted line intensities from all spatial grid points and 
used the width of this distribution to estimate uncertainties and significances,
for both $\gamma$-ray lines of interest (see Fig.~\ref{Georgii:Line}).
The exposure and effective detector area were used to convert line intensities
and uncertainties into source photon flux units. 

No significant difference between the spectra of the supernova and off-supernova 
positions was found. 
We obtained a $2 \sigma$ upper flux limit of $4.1 \cdot 10^{-5}$ photons~cm$^{-2}$~s$^{-1}$ 
for the 847 keV line and $2.3 \cdot 10^{-5}$ photons~cm$^{-2}$~s$^{-1}$ for the 1238 keV 
line.

\subsection{Imaging analysis}
In this analysis the imaging capability of COMPTEL was fully exploited, thus 
increasing the sensitivity of the instrument mainly in the 847 keV line due to a 
better background suppression. 
In { the ``low-threshold'' observing mode}
most of the D2 detector thresholds are well below the 847 keV line of interest.
However, there is a considerable spread of thresholds, the lowest
value being 450~keV.
In standard COMPTEL analysis, event selection with D2 energy deposits above 
650~keV is applied, to simplify the modular
composition of the instrument response function. This is now inappropriate, and
response function composition and usage is more complex, as a price for the
improved 847 keV sensitivity.
For the upper detector plane D1, all modules have hardware thresholds
in a narrow range at or below 50 keV, thus still allowing the use of a homogeneous 
analysis threshold at 50~keV for all D1 detectors.

\begin{figure*}[tbh]
\centering
\begin{minipage}[t]{8.5cm}
\includegraphics[width=\textwidth,bb= 45 45 440 250,clip]{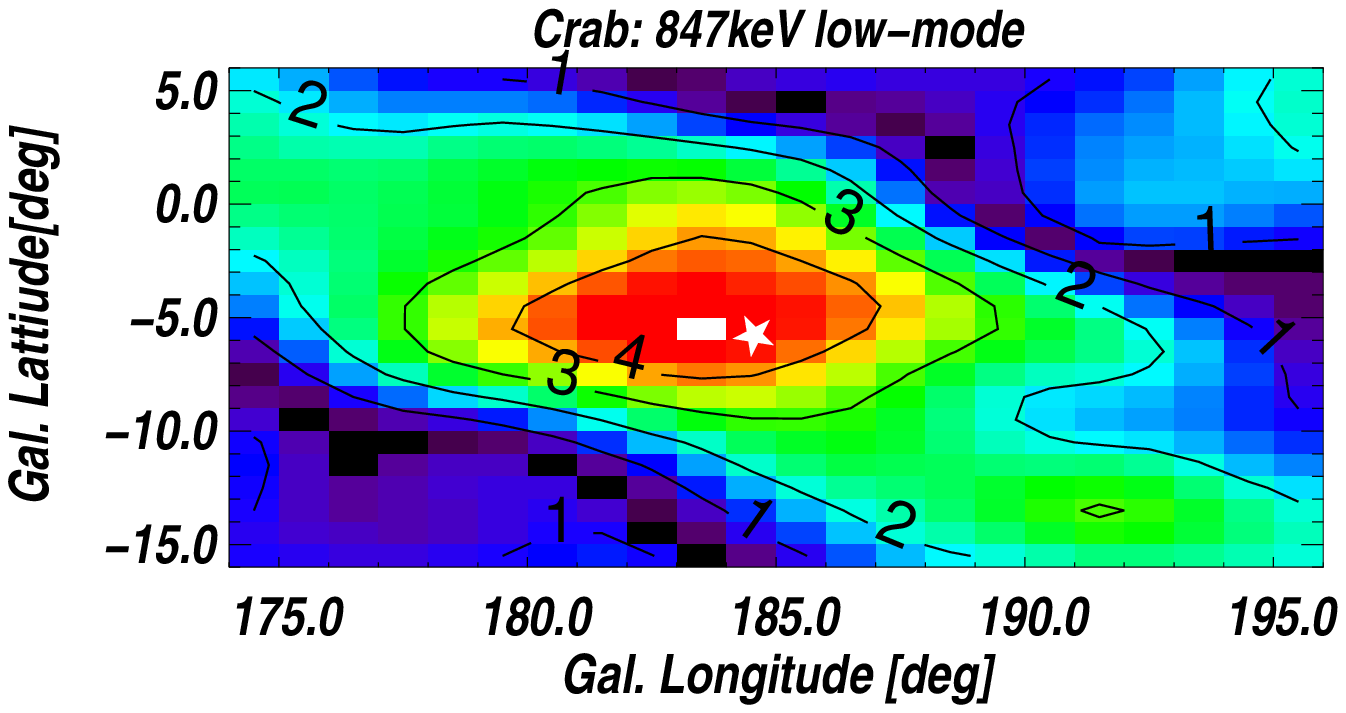}
\end{minipage}
\hfill
\begin{minipage}[t]{8.5cm}
\includegraphics[width=\textwidth,bb= 45 45 440 250,clip]{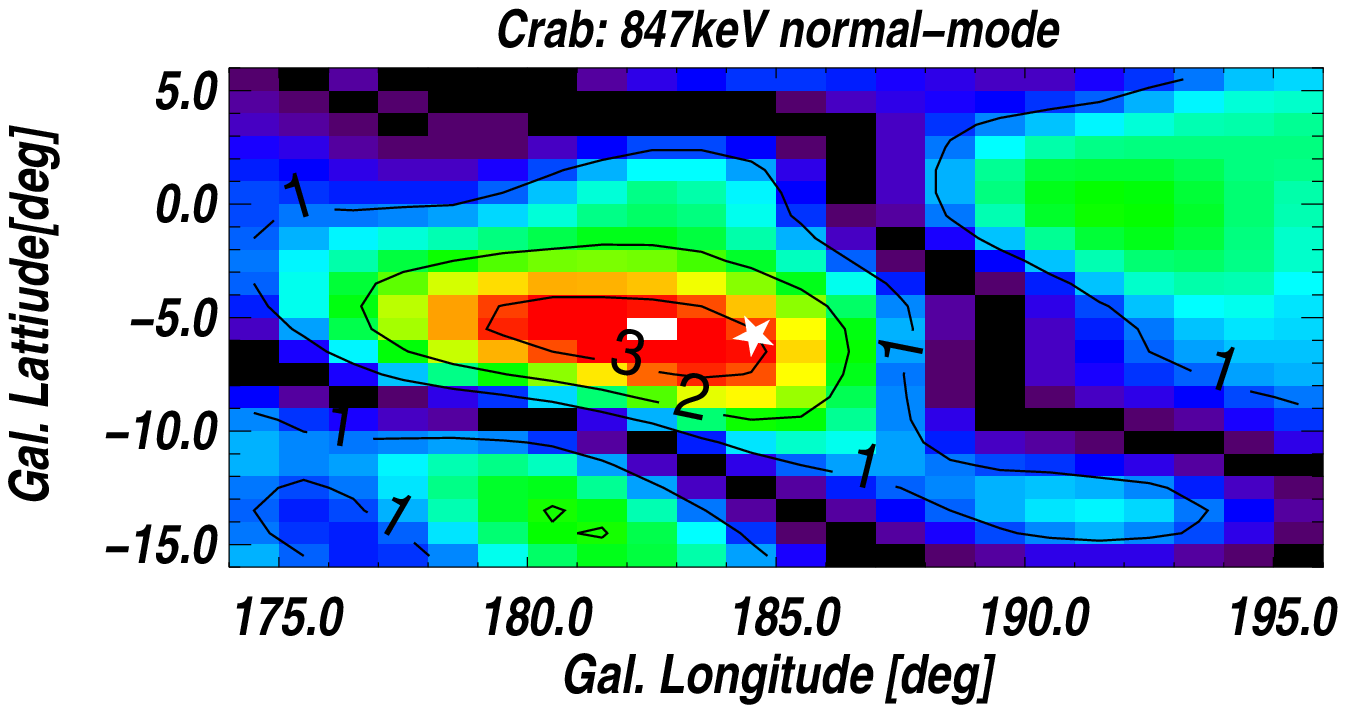}
\end{minipage}
\caption{Improvement of the sensitivity with the low-threshold mode response function (left) for 3 day of Crab data  compared
 to the sensitivity with the standard response function (right). Shown is a significance map based on a maximum-likelihood method.
 $\sigma$ contour lines are plotted. The position of Crab is marked with a star.}
\label{Georgii:F11}
\end{figure*}
\begin{figure*}[tph]
\centering
\includegraphics[width=17cm,bb= 0 0 670 510,clip]{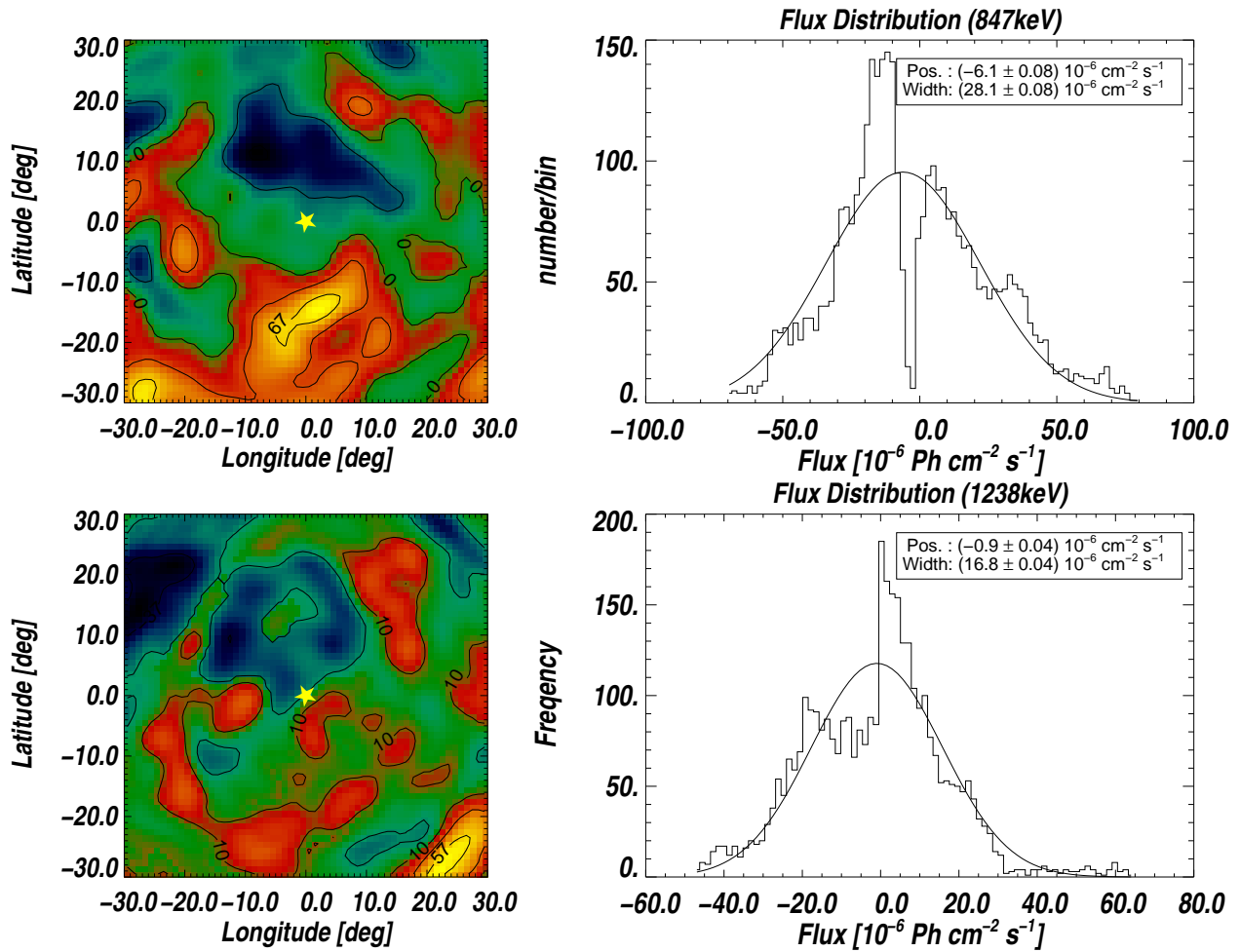}
\caption{Flux maps of the SN1998bu region in 847 keV (top left) and 1238 keV (bottom left). 
No significant excess is seen around the position of the SN, which is marked with a star. 
The contour levels are in units of $10^{-6}$ photons~cm$^{-2}$~s$^{-1}$. Right to each map its flux 
distribution and a Gaussian fit to it is shown. The distortion of the distributions around small 
flux values is due to the method applied.}
\label{Georgii:F3}
\end{figure*}
\begin{figure*}[htp]
\sidecaption
\includegraphics[height=12cm,angle=90]{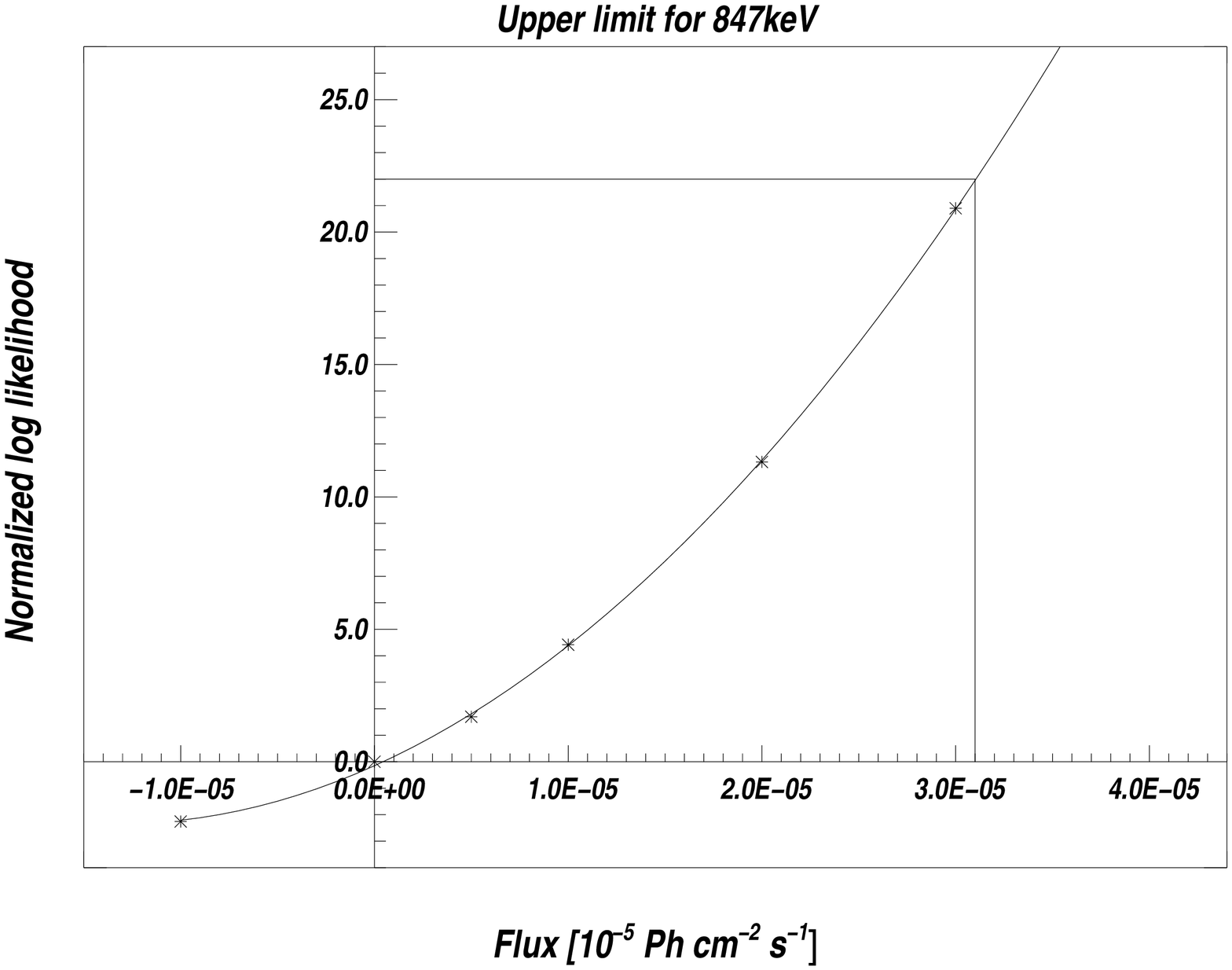}
\caption{The flux limit for the 847 keV line, derived from combined imaging 
results per each minitelescopes. 
We fix a source flux at the position of the SN, and determine the
maximum log-likelihood ratio value for each detector subset. We can 
add these, as derived from independent data. We normalize to the 
log-likelihood ratio of our best-fit flux value of 
$0$ photons~cm$^{-2}$~s$^{-1}$. 
Varying adopted fluxes, we derive the parabolic behavior near the 
minimum (see stars). For our 11 independently-varied parameters 
(one per each detector subset) we obtain a 2$\sigma$ SN flux limit 
by increasing the likelihood by 22 units above the minimum 
(i.e., $3.1 \cdot 10^{-5}$photons~~cm$^{-2}$~s$^{-1}$).} 
\label{Georgii:F4}
\end{figure*}

Instrument imaging response functions were derived for our analysis from
simulations, where each D2 module was treated with its effective hardware threshold. 
The full-instrument data were analyzed in two 
different ways: We first used a summed response function for both lines, but for the 847 keV line
we also analyzed separately for 
``mini-telescopes'' and composed the individual results.

We demonstrate the difference between standard analysis and our more complex
``low-energy-threshold-mode'' response function on 
3 days of Crab low-threshold data, fortuitously collected during an observation of
Geminga. Using the low-mode response we clearly detect Crab with 4.5$\sigma$
at 847 keV, compared to only 3.0$\sigma$ with the
standard response (see Fig.~\ref{Georgii:F11}). For the 1238 keV line the
values are 1.6$\sigma$
and 0.63$\sigma$, respectively. This 
shows that most D2 modules are now sensitive below the
650 keV analysis threshold used for the standard analysis,
resulting in a larger sensitivity mainly for the 847 keV line.

Unfortunately, due to the strong background line at 511~keV, 
we did not obtain the theoretically-achievable sensitivity at the 847 keV line.
To reduce the impact of this background via its scattering pattern, 
we applied a cut of 40$^{0}$ in the angle between 
the incident and the scattered  $\gamma$-ray (the so-called
$\bar{\varphi}$ angle \cite[see][]{Georgii:Schoenfelder1993}). This sacrifices a fraction of source events.

Using these simulated summed responses and the $\bar{\varphi}$ cut, 
maps for the two  $\gamma$-ray energies (shown on the left side in Figure
\ref{Georgii:F3}) are produced through a maximum-likelihood method.
There is no detectable signal from the supernova.

For determination of an upper limit on the flux in the $\gamma$-ray lines we estimated 
statistical and systematic uncertainties. In a histogram of flux values for all pixels 
in the maps of Fig.~\ref{Georgii:F3}, 
the flux variances provide a global measure of the uncertainty.  
Using the Bayesian method described in
\cite{Georgii:Georgii1997}, which accounts for the systematic
and statistical uncertainties, 2$\sigma$ upper limits of $5.5
\cdot 10^{-5}$ photons~cm$^{-2}$~s$^{-1}$ for the 847 keV line and of
$3.2 \cdot 10^{-5}$ photons~cm$^{-2}$~s$^{-1}$ for the 1238 keV line were determined. 
The less constraining limit for the 847 keV line is a result of the $\bar{\varphi}$ cut for background
suppression.

\begin{figure*}[htp]
\centering
\begin{minipage}[t]{8.5cm}
\includegraphics[width=\textwidth,bb= 25 10 440 280,clip]{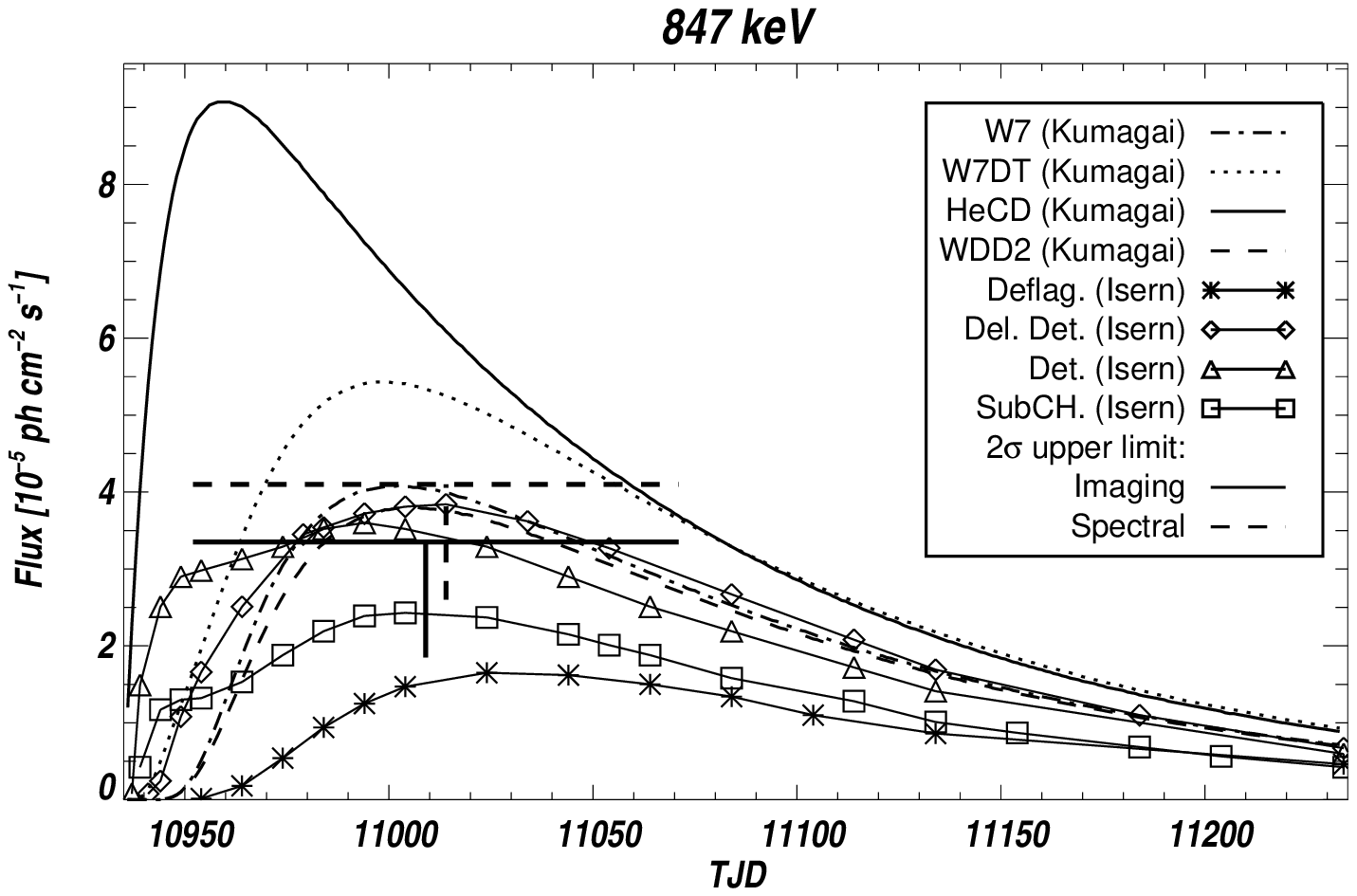}
\end{minipage}
\hfill
\begin{minipage}[t]{8.5cm}
\includegraphics[width=\textwidth,bb= 25 10 440 280,clip]{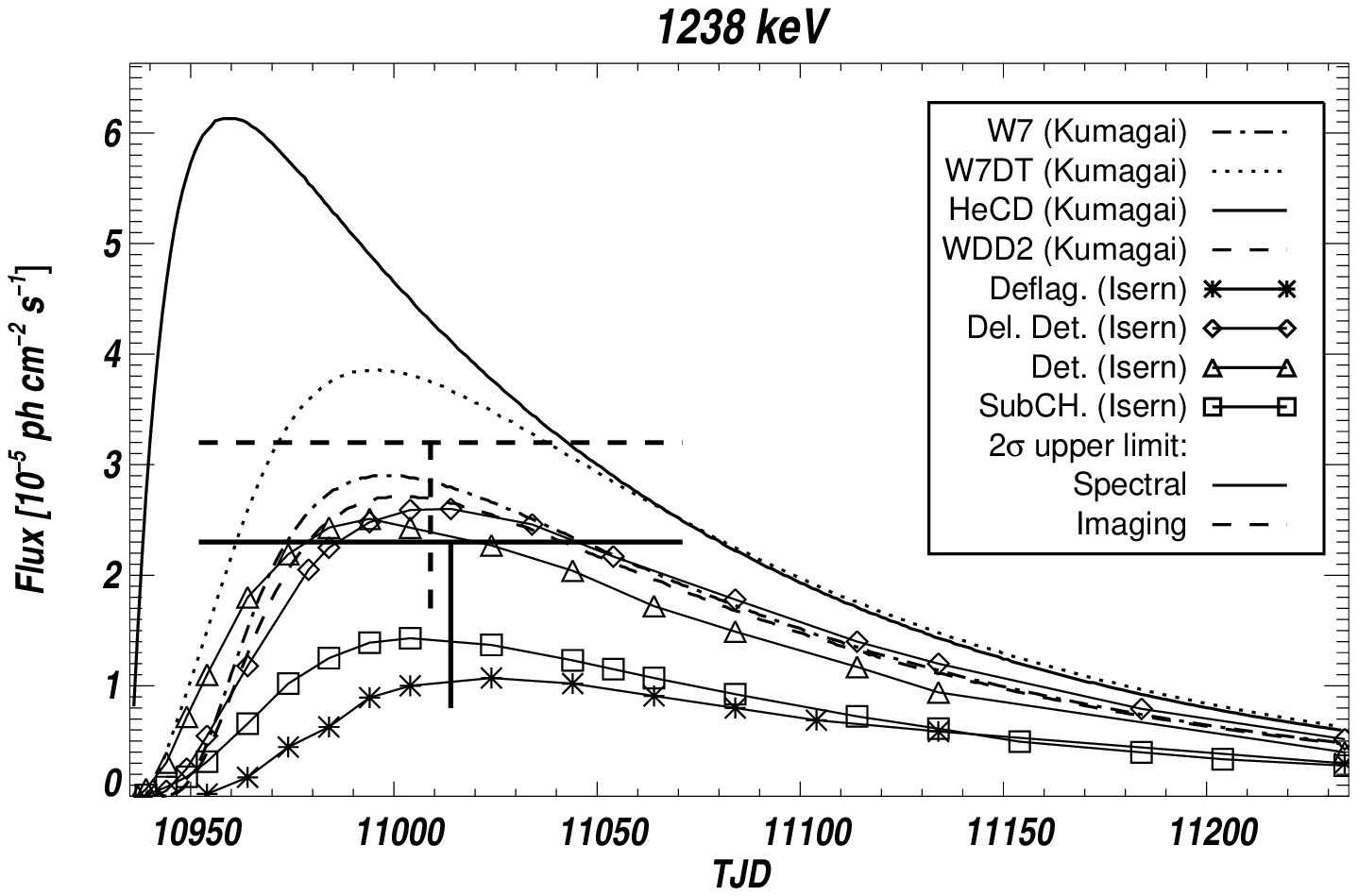}
\end{minipage}
\caption{The model fluxes for the 847 keV (left) and for the 1238 keV (right) line for different  models versus
time after the explosion for a distance of 11.3 Mpc. The upper limits from the spectral and imaging analysis 
are also shown. The solid line represents the more sensitive value from both methods for each line. Note that 
for the 847 keV line the imaging analysis and for the 1238 keV line the spectral analysis is more sensitive.}
\label{Georgii:F5}
\end{figure*}

Therefore another analysis method was applied to the 847 keV line, only.
Here separate maximum-likelihood imaging of the SN1998bu region was made for each mini-telescope 
(the combination of one single D2 module with all D1 modules) by applying the specific response for each D2 module.  
With energy cuts at 650~keV, but no $\bar{\varphi}$ cut, 511 keV background suppression retains better 847~keV sensitivity than above. Since the likelihood values are additive (probabilities) we combined these results
to derive a 2$\sigma$ upper limit of $3.1 \cdot 10^{-5}$ photons~cm$^{-2}$~s$^{-1}$ (see Fig.~\ref{Georgii:F4}). 

\subsection{Summarizing the results} From the imaging analysis we obtain a  2$\sigma$ upper limit of
$3.1 \cdot 10^{-5}$ photons~cm$^{-2}$~s$^{-1}$ for the 847 keV line and $3.2 \cdot 10^{-5}$ photons~cm$^{-2}$~s$^{-1}$
for the 1238 keV line. Spectral analysis yields 
$4.1 \cdot 10^{-5}$ photons~cm$^{-2}$~s$^{-1}$ for the 847 keV line and $2.3 \cdot 10^{-5}$ photons~cm$^{-2}$~s$^{-1}$ 
for the 1238 keV line, respectively. The reason for the different values for both lines in the spectral 
analysis lies in the steep decrease of the background with the energy. In contrast, in imaging 
analysis the background on the sky around the source is not much different for both lines. 

The analysis presented here was based on the assumption that the emitted lines are narrow. This is 
justified because of the energy resolution of COMPTEL (about 88 keV for the 847 keV and about 110 keV 
for the 1238 keV line). However, due to the velocity of the expelled matter, the lines are broadened 
and even shifted (only shortly after the explosion) by the Doppler effect 
(see \cite{Georgii:Hoeflich1998,Georgii:Isern1997}). An analysis of the expected degradation of the upper 
limits yields for a persistent Doppler broadening of the 1238 keV line of 10 000 km/sec \citep{Georgii:Hoeflich1998}, 
an increase of the $2\sigma$ upper limit flux by 12 \% compared to the narrow-line assumption. 
This changes the flux upper limits only marginally and we therefore neglected it.

\section{Discussion}
To compare  $\gamma$-ray upper limits with theoretical Ni mass predictions, the distance to the SN
plays an essential role. The host galaxy M96 had
a HST-Cepheid-determined distance of 11.6 $\pm$ 0.9 Mpc
\citep{Georgii:Tanvir1995}, later revised to 11.3 $\pm$ 0.9 Mpc by
\cite{Georgii:Hjorth1997}. This makes SN1998bu one of seven SN
observed in galaxies with a distance set by Cepheid measurements. We adopt this distance of 11.3 Mpc for our analysis.

A second key factor is the transparency of the supernova to $\gamma$-rays. This is a key issue
determining how much radioactive energy is converted into kinetic energy and supernova light.
The maximum of the optical light curve (about 10 days after the explosion) is set by a maximum of
the product of (declining) energy deposition and (rising) $\gamma$-ray energy escape \citep{Georgii:PintoE2001}.
Furthermore, the Compton scattering optical depth to 1 MeV $\gamma$-rays is below unity beyond about 50 days
after explosion \citep{Georgii:Pinto2001}, but absorption corrections to observed line $\gamma$-rays 
are probably significant for typical models up to about 100 days after explosion 
\citep[see e.g. Fig. 11 in][calculated for energies down to 10~keV, however]{Georgii:Hoeflich1998}. 
This illustrates clearly the importance of $\gamma$-ray measurements with high spectral resolution and
at those early times, in order to directly address the explosion mechanism: the line shapes and the
ratio of the different line intensities from the $^{56}$Ni decay chain can reveal the ratio between
deposited and directly radiated radioactive energy (e.g. \cite{Georgii:Hoeflich1996}). 

For illustrative purposes and simplification, we may simply assume as an extreme case that the 
supernova was transparent for our observation of the
gamma-rays from $^{56}$Co decay over days 17-136 with emphasis on the late part; 
in this case we directly convert our flux
limits into $^{56}$Co (and therefore original $^{56}$Ni) masses. 
Our lowest upper limit for the 1238 keV line of 
$2.3 \cdot 10^{-5}$ photons~cm$^{-2}$~s$^{-1}$ then constrains the visible $^{56}$Ni mass to below 0.35 \Msol.
If we then want to reconcile this with the 0.77 \Msol\ of total $^{56}$Ni determined bolometrically \citep{Georgii:Leibundgut2000}, 
more than half of the $\gamma$-ray energy would be deposited in the supernova over this time window.
We therefore do have to look in detail at the energy deposition efficiency around peak optical luminosity and/or 
effectiveness of $\gamma$-ray escape soon thereafter. 

For several model classes (detonation, delayed detonation, and sub-Chandrasekhar), $\gamma$-ray light curves have been 
calculated through detailed Monte-Carlo photon transport in the expanding supernova 
\citep{Georgii:Hoeflich1998,Georgii:Kumagai1998,Georgii:Isern1997,Georgii:Pinto2001}. Considerable variety in the
gamma-ray flux by factors up to 5 arises from the different explosion models, envelope structures, and photon transport
treatments employed in such calculations. 
In Fig.~\ref{Georgii:F5}, expected  $\gamma$-ray light curves for a few typical models 
\citep{Georgii:Isern1998,Georgii:Kumagai1998} 
are shown, re-scaled for a distance of 11.3 Mpc.
In Table  \ref{Georgii:T1} we list the $^{56}$Ni mass for each of these models, together with time-averaged fluxes 
over the observation time for each $\gamma$-ray line. We see that the predicted $^{56}$Co $\gamma$-ray flux
does not follow the straightforward scaling to the amount of $^{56}$Ni, the explosion mechanism and
the envelope photon transport determine the time-dependent $\gamma$-ray fluxes. 
For the same type of explosion model, predicted $^{56}$Ni masses vary within a factor of two: For the delayed-detonation
class of models, values between 0.55 \Msol\ and 0.96 \Msol\ have been published 
\citep{Georgii:Iwamoto1999,Georgii:Woosley1991,Georgii:Isern1997}, as a result of differences in the
point at which the initially-slow nuclear burning (deflagration) is assumed to turn into a detonation.  
This typical intrinsic variability within an explosion type of a factor of two, which
directly translates into the   $\gamma$-ray flux scaling, 
indicates the systematics which typically remains, within an explosion type. 
In fact, any of the SNe~Ia scenarios (sub-Chandrasekhar, deflagration,
delayed detonations, and pulsating delayed detonation models) has been shown to be capabable to
produce a wide variety of $^{56}$Ni masses ranging from $\simeq$~0.1 to 1 \Msol\ 
(e.g. \cite{Georgii:Nomoto1984,Georgii:Hoeflich1996,Georgii:Hoeflich2002}).

On the other hand, Fig.~\ref{Georgii:F5} illustrates that, for approximately the same amount of total $^{56}$Ni, 
pure deflagration or detonation models are about a factor of two dimmer in   $\gamma$-rays, while the sub-Chandrasekhar
model with substantial $^{56}$Ni sitting further outside reaches a   $\gamma$-ray flux about twice as large as
the typical delayed-detonation model.

\begin{table}[h]
\caption{Fluxes in the 847 keV 
and 1238 keV line for different models, averaged over our observation time.}
\label{Georgii:T1}
\begin{minipage}{\textwidth}
\begin{tabular}{llll}
\hline
Model & $^{56}$Ni mass & I$_{\rm 847~keV}$$\cdot$10$^5$ & I$_{\rm 1238~keV}$$\cdot$10$^5$ \\
&[\Msol] & [ph~cm$^{-2}$s$^{-1}$]&[ph~cm$^{-2}$s$^{-1}$]\\ \hline
        W7\footnote{\cite{Georgii:Kumagai1998}}   & 0.58 & 4.2&3.0\\
        W7DT$^a$                                  & 0.77 & 5.8&4.1\\
        HeCD$^a$                                  & 0.72 & 8.2&5.5\\
        WDD2$^a$                                  & 0.58 & 3.9&2.8\\
        Deflag.\footnote{\cite{Georgii:Isern1998}}& 0.50 & 1.5&1.0\\
        Del. Det.$^b$                             & 0.80 & 4.3&2.7\\
        Det.$^b$                                  & 0.70 & 4.0&2.7\\
        SubCH.$^b$                                & 0.60 & 2.7&1.5\\ \hline
\end{tabular}
\end{minipage}
\end{table}

In Table  \ref{Georgii:T1} we also list the time-averaged $^{56}$Co   $\gamma$-ray fluxes of each model,  
together with our 2$\sigma$ upper limits.  
Our SN1998bu flux limits are well below  the "HeCD"
and the "W7DT" model predictions for both lines. 
The  "W7", "WDD2", delayed detonation and detonation model fluxes are marginally 
consistent with our flux limits, while the fluxes predicted from the sub-Chandrasekhar model 
and the deflagration model are 
consistent with our limits for both line energies at the adopted distance. 

This comparison illustrates that with a $^{56}$Ni mass in the "typical" range derived for SN1998bu, around
0.7-0.8 \Msol\ \citep{Georgii:Leibundgut2000}, we should have seen $^{56}$Co   $\gamma$-rays at least due to those models
which turn more rapidly from deflagration into detonation (W7DT) or partially-produce radioactivity in their outer
ejecta (HECD). Yet, within Chandrasekhar-type models of the presently-favored type of a delayed
transition from deflagration into detonation, a total $^{56}$Ni mass as high as
about 1 \Msol\ may still be consistent with our measurement. 

At a distance of 9.6 $\pm$ 0.6 Mpc \citep{Georgii:Feldmeier1997} based on planetary nebulae (PN) all 
models would be inconsistent with our 1238 keV and 847 keV flux limits, indicating either this distance 
is incorrect (see \citep{Georgii:Maoz1999} for a discussion on a possible correction in the distance 
ladder scale) or that model treatments generally overestimate the $^{56}$Ni masses.

It will require time-resolved measurements of the   $\gamma$-ray flux (hence a brighter / more nearby supernova or a
more sensitive instrument), or exploitation of spectral-shape details as promised by the spectrometer aboard INTEGRAL
\citep[see discussion in][]{Georgii:Isern1997}, to decide among explosion models from
gamma-ray line measurements alone.


\begin{acknowledgements}
The COMPTEL project is supported by the German government through DLR grant 50 QV 90968, by 
NASA under contract NAS5-26645, and by the Netherlands Organization for Scientific Research (NWO).
\end{acknowledgements}


\begin{thebibliography}{}
\bibitem[Ayani et~al.(1998)]{Georgii:Ayani98}Ayani, K., Nakatani, H., Yamaoka, H., IAU Circ. No, {\bf 6905}, 1998.
\bibitem[Branch(1998)]{Georgii:Branch1998}Branch D., ARA\&A, {\bf 36}, 17--55, 1998.
\bibitem[Bravo et al.(1993)]{Georgii:Bravo1993}Bravo, E., Dominguez, I.,  Isern, J, et al., A\&A {\bf 269}, 187--194, 1993. 
\bibitem[Cappellaro et al.(2001)]{Georgii:Cappellaro2001}Cappellaro, E., Patat, F., Mazzali, P.~A., et al., ApJ {\bf 549}, L215--L218, 2001.
\bibitem[Contardo et al.(2001)]{Georgii:Contardo2000}Contardo G., Leibundgut B., Vacca W.D., A\&A {\bf 359}, 876--886, 2000.
\bibitem[Faranda \& Skiff(1998)]{Georgii:Faranda1998}Faranda ,C. and Skiff, B.A., IAU Circ. No. {\bf 6905}, 1998.
\bibitem[Feldmeier et~al.(1997)]{Georgii:Feldmeier1997}Feldmeier, J.J., Ciardullo, R. and Jacoby, G.H., ApJ {\bf 479}, 231--243, 1997.
\bibitem[Georgii et~al.(1997)]{Georgii:Georgii1997}Georgii, R., Diehl, R., Lichti, G., et al., Proc. 2$^{nd}$ INTEGRAL workshop, ESA {\bf SP-382}, 51--54, 1997.
\bibitem[Hjorth \& Tanvir(1997)]{Georgii:Hjorth1997}Hjorth, J. and Tanvir, N.R., ApJ {\bf 482}, 68--74, 1997.
\bibitem[H\"oflich et~al.(1998)]{Georgii:Hoeflich2002}H\"{o}flich P., Gerardy C. L., Fesen R. A., Sakai S., ApJ {\bf 568}, 791, 2002
\bibitem[H\"oflich et~al.(1998)]{Georgii:Hoeflich1998}H\"{o}flich, P., Wheeler, J.C. and Khokhlov, A.,
         ApJ {\bf 492}, 228--245, 1998.
\bibitem[H\"oflich \& Khoklov(1996)]{Georgii:Hoeflich1996}H\"{o}flich, P., and Khokhlov, A.,
         ApJ {\bf 457}, 500, 1996.
\bibitem[Iben \& Tutukov(1984)]{Georgii:Iben1984}Iben I. Jr., and Tutukov A.V., ApJS {\bf 54}, 335, 1984.
\bibitem[Isern(1997)]{Georgii:Isern1997}Isern, J., G\`omez-Gomar J., Bravo E., Jean P., ESA-SP-382, 89, 1997
\bibitem[Isern(1998)]{Georgii:Isern1998}Isern, J., private communication, based on \citep{Georgii:Isern1997,Georgii:Bravo1993}
\bibitem[Iwamoto et~al.(1999)]{Georgii:Iwamoto1999}Iwamoto,K., Brachwitz, F., Nomoto, K., et al., ApJS {\bf 125}, 439--462, 1999.
\bibitem[Jah et al.(1999)]{Georgii:Jah1999}Jha,S. ,Garnavich, P.M., Kirshner, R.P., et al., ApJS {\bf 125}, 73--79, 1999.
\bibitem[Leibundgut(2000)]{Georgii:Leibundgut2000} Leibundgut B., Astr.Astroph.Rev. {\bf 10}, 179-209, 2000.
\bibitem[Livio(2000)]{Georgii:Livio2000}Livio M., in "Type Ia Supernovae: Theory and Cosmology", eds. Niemeyer J.C. and Truran J.W., Cambr. Univ. Press, p 33--48, 2000.
\bibitem[Livne(1990)]{Georgii:Livne1990}Livne E., ApJ {\bf 354}, L53--L55, 1990.
\bibitem[Livne(1995)]{Georgii:Livne1995}Livne E., Arnett D., ApJ {\bf 452}, 62--74, 1995.
\bibitem[Kumagai(1998)]{Georgii:Kumagai1998}Kumagai, S., private communication, based on \citep{Georgii:Kumagai1997}.
\bibitem[Kumagai and Nomoto(1997)]{Georgii:Kumagai1997}Kumagai S. and Nomoto K., in "Thermonuclear Supernovae", eds. R. Canal, P. Ruiz-Lapuente and J. Isern, p 515.
\bibitem[Maoz et~al.(1999)]{Georgii:Maoz1999}Maoz, E., Newman, J.A., Ferrarese, L., et al., Nat. {\bf 401}, 351--354, 1999.
\bibitem[Meikle et~al.(1998)]{Georgii:Meikle1998}Meikle, P., Hernandez, M., Fassia, A., IAU Circ. No, {\bf 6905}, 1998.
\bibitem[Morris et~al.(1997)]{Georgii:Morris1997}Morris, D.J., Bennett, K., Bloemen, H., et al., Proc. of the 4$^{th}$ Compton Symposium, AIP Conf. Proc. {\bf 410}, 1084--1088, 1997.
\bibitem[Niemeyer \& Truran (2000)]{Georgii:Niemeyer2000}Niemeyer J.C. and Truran J.W. (eds.), "Type Ia Supernovae: Theory and Cosmology", Cambr. Univ. Press, 2000.
\bibitem[Nomoto et~al.(1997)]{Georgii:Nomoto97}Nomoto, K., Iwamoto, K., Nakasato, N., et al., Nucl. Phys. {\bf A621}, 467--476, 1997.
\bibitem[Nomoto et~al.(1984)]{Georgii:Nomoto1984}Nomoto, K., Thielemann F.-K., Yokoi K., ApJ {\bf 286}, 644--658, 1984.
\bibitem[Nomoto(1982)]{Georgii:Nomoto1982}Nomoto, K., ApJ {\bf 253}, 798--810, 1982.
\bibitem[Pinto et~al.(2001)]{Georgii:Pinto2001}Pinto, P.A., Eastman, R.G. and Rogers, T., ApJ. {\bf 551}, 231--243, 2001.
\bibitem[Pinto and Eastman(2001)]{Georgii:PintoE2001}Pinto, P.A., and Eastman, R.G., T., New Astr. {\bf 6}, 307--319, 2001.
\bibitem[Reinecke et~al.(1999)]{Georgii:Reinecke1999}Reinecke M., Hillebrandt W., Niemeyer J.C., A\&A {\bf 347}, 739--755, 1999.
\bibitem[Sch\"onfelder et~al.(1993)]{Georgii:Schoenfelder1993}Sch\"{o}nfelder, V., Aarts, H., Bennett, K., et al.,
         ApJS {\bf 86}, 657--692, 1993.
\bibitem[Tanvir et~al.(1995)]{Georgii:Tanvir1995}Tanvir, N.R., Shanks, T., Ferguson, H.C., Robinson, D.R.T.,
         Nature {\bf 377}, 27--31, 1995.
\bibitem[Villi(1998)]{Georgii:Villi98}Villi, M., IAU Circ. No. {\bf 6899}, 1998.
\bibitem[Vinko et~al.(2001)]{Georgii:Vinko2001}Vinko, J., Csak, B., Csizmadia, Sz., et al., A\&A {\bf 372}, 824--832, 2001.
\bibitem[Weidenspointner et~al.(2001)]{Georgii:Weidenspointner2001} Weidenspointner, G., Varendorff, M., Oberlack, U., et al., A\&A {\bf 368}, 347--368, 2001.
\bibitem[Woosley and Weaver(1991)]{Georgii:Woosley1991} Woosley S.E., Weaver T.A., Les Houches Lecture Notes, eds. J. Audouze et al. (Amsterdam:Elsevier), 1991.
\end{thebibliography}
\end{document}